# Finite size effect on Néel temperature with $Co_3O_4$ nanoparticles

Lin He, Chinping Chen[a]*

Department of Physics, Peking University, Beijing, 100871, People's Republic of China

Ning Wang, Wei Zhou, Lin Guo[b]

School of Materials Science and Engineering, Beijing University of Aeronautics and Astronautics, Beijing, 100083, People's Republic of China

Abstract:

Finite size effect on the antiferromagnetic (AFM) transition temperature, $T_N$, has been investigated with $Co_3O_4$ nanoparticles of 75, 35, and 16 nm in diameter. Along with the results from the previous experiments on the $Co_3O_4$ nanoparticles with average diameter of 8 and 4.3 nm, the variation of $T_N$ with the particle diameter, $d$, appears to follow the finite size scaling relation. The shift exponent is determined as $\lambda = 1.1 \pm 0.2$, and the correlation length, $\xi_0 = 2.8 \pm 0.3$ nm. The value of the shift exponent is consistent with the description by the mean field theory.

## I. Introduction

Recently, the properties of magnetic nanoparticles have received much attention due to the interesting and fruitful phenomena compared to that of bulk phase. In particular, the properties of the AFM $Co_3O_4$ nanoparticles are one of the focused points. For example, freezing temperature of 8.4 K has been reported with porous nanotubes of $Co_3O_4$ with the tube diameter ranging from a few tens to 200 nm [1], exchange biased effect arising from the AFM core and surface spins has been observed with 8 nm $Co_3O_4$ nanowires [2]. The bulk $Co_3O_4$ has a cubic spinel structure with the lattice constant, a = 0.8065 nm. The $Co^{3+}$ ions at the octahedral sites (B sites) are diamagnetic while the spin moments of the $Co^{2+}$ ions at the tetrahedral sites (A sites) exhibit AFM ordering at $T < T_N$ = 40 K [3-6]. In the nanoscale region, some experiments have reported a reducing $T_N$. For example, $T_N$ ~30 K has been observed with 8 nm $Co_3O_4$ nanoparticles [2,3] and 15 ± 2 K with 4.3 nm nanoparticles [7].

By reducing the particle size, an intrinsic finite size effect causing a reduction on the magnetic ordering temperature occurs. In the region close to the magnetic ordering transition, the correlation length $\xi(t)$ of the fluctuation in the order parameter, $M$, diverges logarithmically with the reduced temperature, $t = (T - T_{order})/T_{order}$, in which the ordering transition temperature, $T_{order}$, is $T_C$ for a ferromagnetic (FM) or ferrimagnetic material and $T_N$, for an AFM material. With the geometric confinement effect on $\xi(t)$, the magnetic transition temperature, $T_N$ or $T_C$, is reduced from the bulk value [8]. The finite size scaling relation gives the fraction of downshift in $T_N$ (or $T_C$) from the bulk value, $T_N(\infty)$ (or $T_C(\infty)$), due to the geometric confinement effect by the size, $d$ [9],

$$\Delta T = \frac{T_N(\infty) - T_N(d)}{T_N(\infty)} = \left(\frac{\xi_0}{d}\right)^\lambda, \qquad (1)$$

in which $\xi_0$ is a constant for the correlation length of bulk phase at a temperature away from the ordering temperature, and $\lambda$ is the shift exponent. For any magnetic system with a reduced dimensionality, for example, a quasi-two dimensional (2D) film with a thickness $d$, a quasi-one dimensional (1D) wire with a wire diameter $d$, or a quasi-zero dimensional (0D) particle with a particle diameter $d$, the scaling relation by Eq. (1) applies equally well if the free surface effect is not accounted for. Usually, the value of the shift exponent, $\lambda$, is model-dependent. By the mean field theory, one obtains $\lambda = 1$ [10], the 3D Heisenberg system, $\lambda = 1.4$ [11], and the 3D Ising system, $\lambda = 1.6$ [12]. Particularly, for $\gamma$-Fe$_2$O$_3$ (meghemite) with spinel structure and the free surface effect accounted for by a core-shell model, the correlation length constant and the shift exponent are obtained by a Monte Carlo simulation with the Ising spins as $\xi_0 \sim 2$ lattice constants and $\lambda \sim 2$ [13].

Experimentally, the intrinsic finite size effect with various magnetic materials has been reported mainly with quasi-2D magnetic films. For example, the thickness dependent magnetic phase transition temperatures, $T_C$ and $T_N$, have been investigated with FM films of Fe [14,15], Co [16], Ni [8] and Gd [8,17,18], and with AFM films of CoO [9,19], Cr [20] and Ho [21]. In comparison, there are much less reports concerning the finite size effect studied on even lower dimensional magnetic samples, such as nanowires or nanoparticles. The finite size effect reported on the Curie temperature, $T_C$, of Ni nanowires [22] is one of the rare examples. This may be ascribed to the difficulty in synthesizing a series of nanoparticles or nanowires with controlled sizes. In addition, for FM nanoparticles, the appearance of superparamagnetism may add difficulty in the determination of $T_C$ [23].

In this paper, we report the fabrication and the observation of the finite size effect on $T_N$ with AFM $Co_3O_4$ nanoparticles of 16, 35 and 75 nm in diameter. The AFM transition temperature, $T_N$, determined in the present study along with the other three results on the $Co_3O_4$ nanoparticles of 8 and 4.3 nm in size reported previously [2,3,7], is found to follow the scaling relation by Eq. (1). According to the present study, the important parameters are determined as $\lambda = 1.1 \pm 0.2$, $\xi_0 = 2.8 \pm 0.3$ nm by fixing the bulk value of the AFM transition point as, $T_N(\infty) = 40$ K [3-6].

## II. Experimental and sample characterization

The three different samples were synthesized by two different chemical processes. The sample of 75 nm has been synthesized as follows. Solutions of 70 mL of 0.4 M ammonium oxalate and 35 mL of 0.6 M $CoCl_2$ were added into 30 mL of pure water by co-current flow at 60 °C. The pH value was adjusted to 8.5 by 5 M aqueous ammonia. All of the three solutions contained 0.2% (wt) of vinylpyrrolidone (PVP; $M$w, 40 000) to prevent the floc from coalescing. After the reaction, the loosened pink product was dried at 30 °C then maintained at 420 °C for 3 hours. The other two small-sized $Co_3O_4$ samples, *i.e.* 35 and 16 nm nanoparticles, were obtained by another process. First, $CoSO_4$ and KOH were dissolved in pure water to form solutions of 1.0 M $CoSO_4$, 4.0 M KOH and 8 M KOH. To form the 35 nm nanoparticles, 15 mL of 4 M KOH solution was slowly dropped into 15 mL of 1.0 M $CoSO_4$ solution. It was then vigorously stirred to obtain precipitate. After the reaction, the loosened pink product was kept with the solution for another 2 hours and then dried at 30 °C and maintained at 420 °C for 3 hours. To obtain the 16 nm nanoparticles, the concentration of the KOH solution was increased from 4.0 to 8.0 M.

The Powder X-ray diffraction (XRD) analysis was performed using a Rigaku Dmax X-ray diffractometer with Zn $K_\alpha$ radiation ($\lambda$ = 0.1541 nm). The morphology and structure of the products were investigated using a S4800 cold field emission scanning electron microscopy (FESEM) and transmission electron microscopy (TEM). The magnetic properties of these three samples have been measured by a MPMS SQUID (Quantum Design) system. Figure 1 is for the XRD patterns of the three samples. It shows a high degree of crystallinity. All of the peaks match well with the Bragg reflections of the standard spinel $Co_3O_4$ structure (Fd3m(227)) [24]. Figure 2 shows typical FESEM images for the samples of 75 and 35 nm along with the TEM image of the 16 nm particle. The 75 nm nanoparticles, as shown in Fig. 2a, are highly aggregated to form columnar structure. Nevertheless, the corpuscular feature for each nanoparticle is apparent. For the 35 and 16 nm nanoparticles shown in Figs. 2b and 2c, the particles are more dispersed spatially due to different synthesizing method.

## III. Measurements and analysis

The temperature dependent magnetization curves shown in Fig. 3 were measured by the zero-field-cooling (ZFC) and field-cooling (FC) modes. To perform the ZFC measurement, as shown by the solid curves, the procedure was to cool the sample under zero applied field down to 5 K, and then applied a small field of 90 Oe for data collection in the warming process. For the FC measurements, as shown by the open symbols, the data were collected in the cooling process by an applied magnetic field of 90 Oe. Both the ZFC and FC curves of the three samples exhibit peak structures, revealing a feature of the AFM phase transition. The inset shows an amplified view for the ZFC curves in the low temperature region. The peaks for the samples of 75, 35 and 16 nm particles are determined by the positions of the maximum, as 39 ± 1 K, 38 ± 1 K and 33 ± 1 K, respectively. The uncertainty is estimated by the step size of measurement in temperature, ~ 1 K. The ZFC and FC curves of the three samples exhibit two additional features along with the peak structure: a) the increase of magnetization with decreasing temperature at low temperature which results in minima around 9 – 11 K with both of the ZFC and FC curves, and b) the discrepancy of the magnetization below $T_N$ by the ZFC and FC measurements. Similar characteristics have also been observed with other spinel AFM nanoparticles, for example, with 16 – 70 nm $CoRh_2O_4$ [25] and with 6.8 – 32 nm $ZnCr_2O_4$ [26]. The increase of magnetization at low temperature below the about K is attributable to the contribution of surface uncompensated spins [25], and the discrepancy of the ZFC and FC curves, to the field induced metastable state of surface frustrated spins during the cooling process [25,26].

To further confirm the AFM phase transition, we carried out the ZFC measurements in various applied fields in addition to $H$ = 90 Oe. Figure 4 shows the ZFC curves of 75 nm particles measured by various applied fields, from $H$ = 50 Oe to 5 kOe. The step size in temperature around the peak is about 1 K. The peak positions are almost independent of the applied field. Although similar peaks also appear in the ZFC measurement for the freezing temperature with a surface spin glass state of the Ni nanochains [27] or for the blocking temperature of the core ferromagnetism with the Ni nanochains [27] and the magnetization reversal with the ferrimagnetic $Mn_3O_4$ nanocubes of 4 nm in size [28], these characteristic temperatures decrease with the increasing applied field during the measurement, which is different from the present temperature dependency of the AFM peak with the $Co_3O_4$ nanoparticles.

The Néel temperatures obtained by the ZFC measurements in various applied fields for these three samples are plotted in Fig. 5. The dashed lines are for the averaged values of $T_N$. They are 39.1 ± 0.2 K, 37.2 ± 0.7 K and 32.3 ± 0.6 K for the samples of 75, 35 and 16 nm particles, respectively. These values are larger than the reported values for the 8 nm $Co_3O_4$ nanoparticles, ~30 K [2,3], and the 4.3 nm $Co_3O_4$ nanoparticles, ~15 ± 2 K [7]. The ordering transition temperature of 75 nm particles is only slightly smaller than that of the bulk $Co_3O_4$ [4-6].

Figure 6 shows $T_N(d)$ versus the particle size, $d$, obtained in the present work along with the other three previously reported results [2,3,7]. These data points are fitted using Eq. (1) with the correlation length constant, $\xi_0$, and the shift exponent, $\lambda$, as fitting parameters and with the bulk transition temperature fixed as $T_N(\infty) = 40$ K. The solid curve in Fig. 6 is for the fitting result. The best fit gives $\lambda = 1.1 \pm 0.2$ and $\xi_0 = 2.8 \pm 0.3$ nm. The inset shows a log-log plot for $\Delta T$ versus $d$. The straight line is plotted according to the fitting result. It indicates that the reduction of $T_N$ with the reducing particles size is attributed to the intrinsic finite size effect.

## V. Discussion

The value of the shift exponent, $\lambda = 1.1 \pm 0.2$, is consistent with the result derived by the mean field theory $\lambda = 1$ [10], and is slightly smaller than the theoretical result $\lambda \sim 1.4$ by the 3D Heisenberg system [11]. Although the finite size effect for $\gamma$-$Fe_2O_3$ nanoparticles with the similar spinel structure as $Co_3O_4$ have been investigated by a Monte Carlo simulation based on the Ising spins, the value of $\lambda = 2$ is much larger than our experimental result [13]. Experimentally, $\lambda \sim 1.55$ is obtained with the AFM CoO film [9,19]. It is close to the result derived by the 3D Ising system due to the large magnetocrystalline anisotropy of the sample, $\sim 2.7 \times 10^7$ J/m$^3$ [29]. In this respect, the magnetocrystalline anisotropy of $Co_3O_4$ is only about $9 \times 10^4$ J/m$^3$ [30]. Therefore, $Co_3O_4$ is expected not to behave like an Ising system. Perhaps, this is one of the reasons to explain the large discrepancy between our experimental result and that of the Monte Carlo simulation based on the Ising spins, even though the investigated system has been modeled using the similar spinel crystal structure. The correlation length, $\xi_0 \sim 2.8$ nm, which is about 3.5 lattice constants of $Co_3O_4$, is also much larger than the value of 1.6 nm, $\sim$2 lattice constants, derived by the Monte Carlo simulation [13]. Quantitative understanding of this discrepancy is an interesting point to study. For a comparison, the correlation length determined for the AFM CoO film [9,19] is about 2 nm, which is equal to 4.7 lattice constants with the lattice constant of 0.426 nm. For the FM material such as the hcp Gd film [8,17,18], $\xi_0$ is about 2.5 nm, equal to 4.3 lattice constants with the lattice constant of 0.577 nm along (0001).

## VI. Conclusion

The finite size effect on the Néel temperatures, $T_N$, for three different sizes of $Co_3O_4$ nanoparticles with diameters of 75, 35, and 16 nm have been investigated. The reduction of $T_N$ with the decreasing particle size follows the finite size scaling relation. The shift exponent $\lambda = 1.1 \pm 0.2$ and the correlation length, $\xi_0 = 2.8 \pm 0.3$ nm, are obtained. The value of the shift exponent determined in the present work is consistent with the theoretically derived value by mean field theory. In addition, the correlation length determined for the $Co_3O_4$ nanoparticles is $\xi_0$ ~2.8 nm, which is about 3.5 lattice constants. In terms of lattice constant, it is smaller than the value of 4.7 lattice constants for a AFM CoO film, or 4.3 lattice constants along (0001) for a FM hcp Gd film.

Acknowledgements

This project was financially supported by National Natural Science Foundation of China (20673009) and Specialized Research Fund for the Doctoral Program of Higher Education (20060006005) as well as by Program for New Century Excellent Talents in University (04-0164).

References

*a) corresponding author : cpchen@pku.edu.cn

b) guolin@buaa.edu.cn

Figures

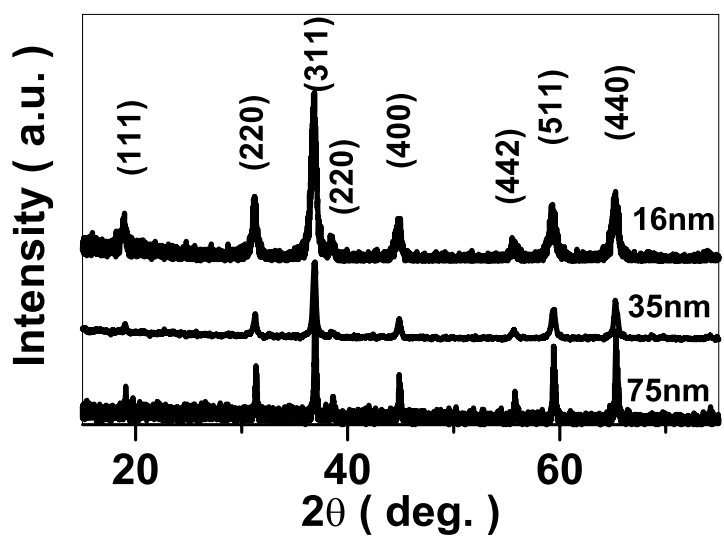

FIG. 1 XRD patterns of the Co$_3$O$_4$ nanoparticles with diameter of 75, 35, and 16 nm.

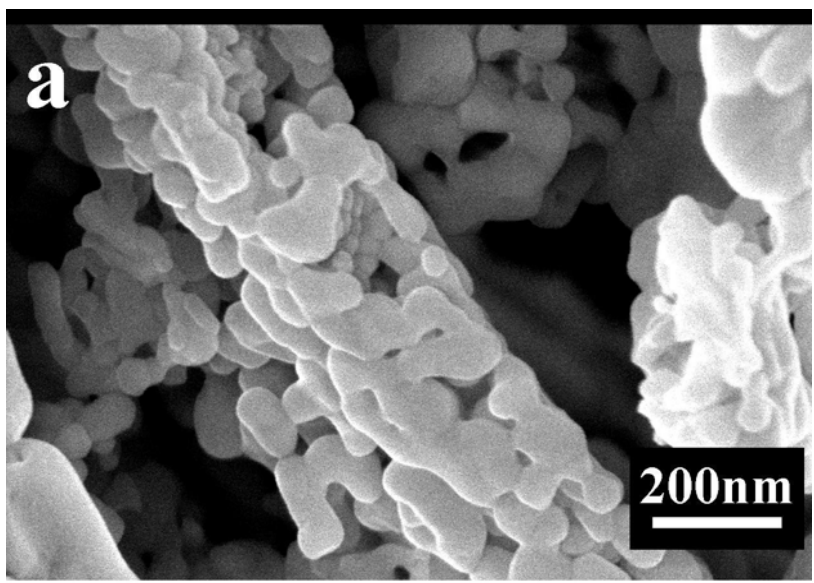

Fig 2a

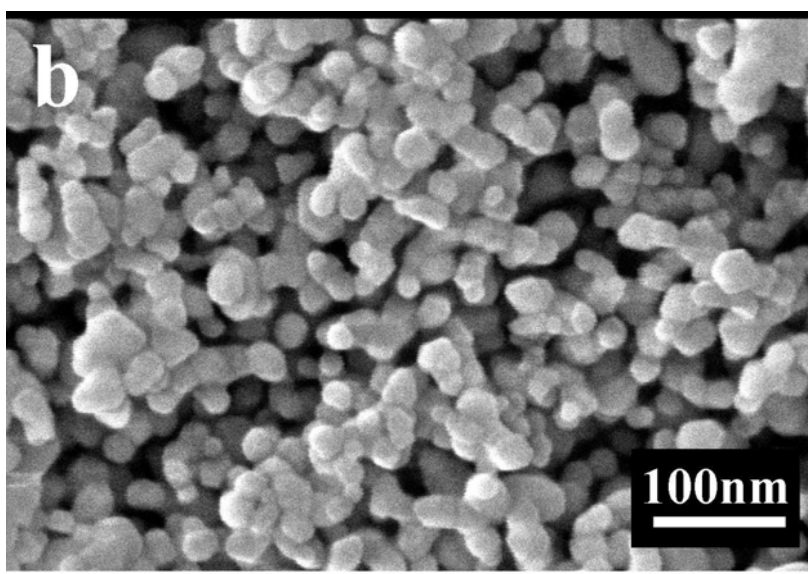

Fig 2b

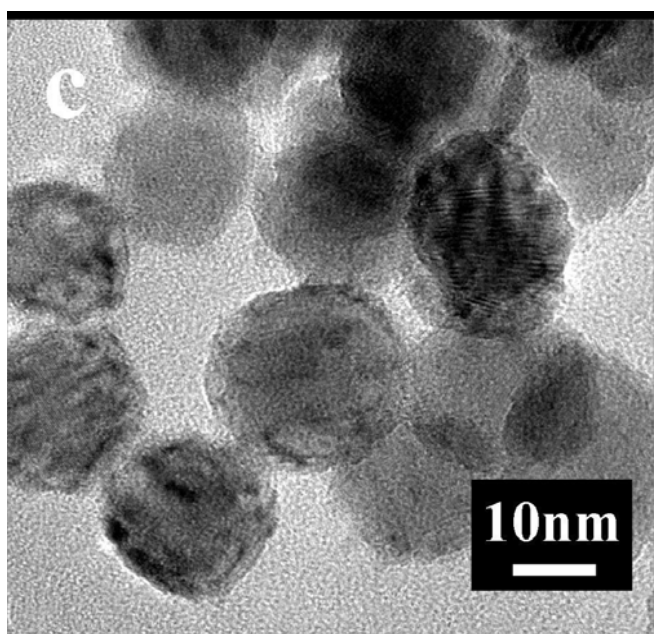

Fig 2c

FIG. 2 FESEM images of the $Co_3O_4$ samples with average diameter of a) 75 nm and b) 35 nm. c) TEM images of the $Co_3O_4$ nanoparticles with average diameter of 16 nm.

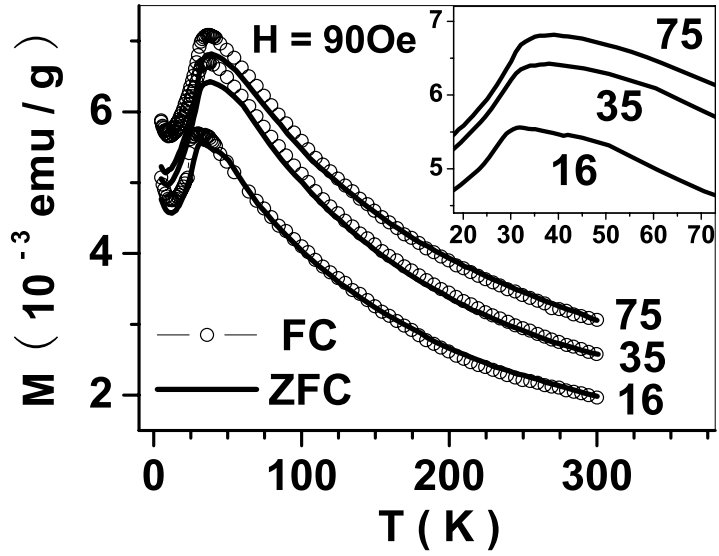

FIG. 3 ZFC (solid lines) and FC (open circles) $M(T)$ measurements for $Co_3O_4$ nanoparticles of 75, 35 and 16 nm in size. The applied field is 90 Oe and the temperature range of measurement is between 5 and 300 K. The inset shows the amplified view around the peaks of the ZFC curves.

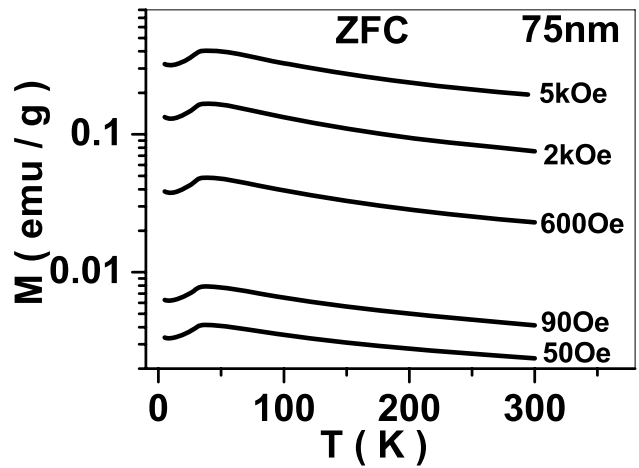

FIG. 4 ZFC curves measured in different applied fields for the $Co_3O_4$ nanoparticles of 75 nm.

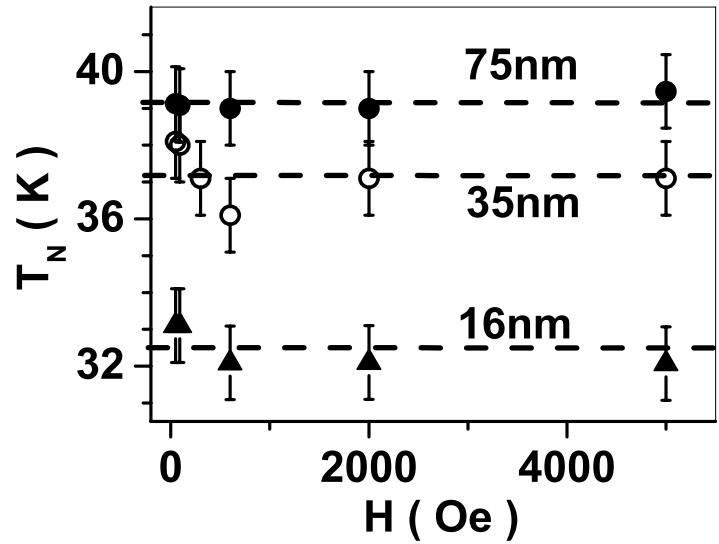

FIG. 5 Néel temperatures determined from the peaks of ZFC curves measured by different applied fields. The dashed lines are the average values of $T_N$ for the 75, 35 and 16 nm nanoparticles.

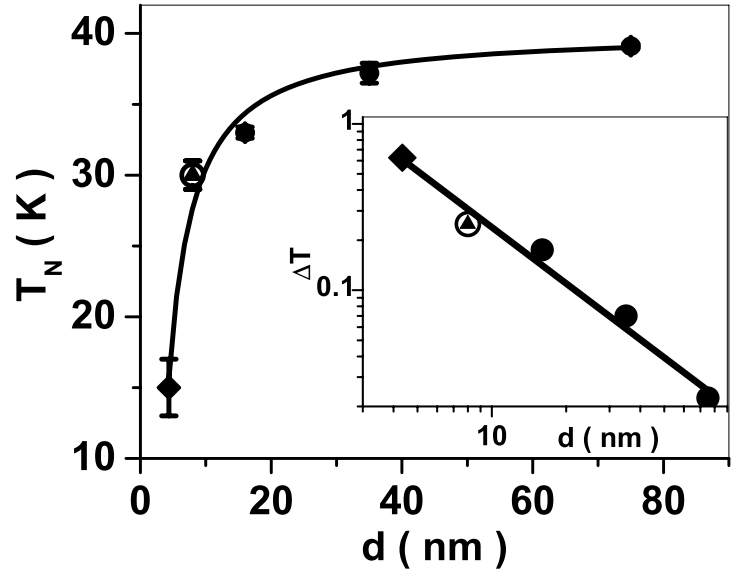

Fig. 6 Néel temperatures, $T_N(d)$, versus diameters, $d$. The solid circles are for the three data points determined in the present work. The open circle and solid triangle are for the published data of $d = 8$ nm from Refs 2 and 3, and the solid diamond are for $d = 4.3$ nm from Ref 7. The solid curve is for the best fit by Eq. (1) with $\lambda = 1.1 \pm 0.2$ and $\xi_0 = 2.8 \pm 0.3$ nm by fixing $T_N(\infty) = 40$ K. The inset shows the log-log plot of the reduced temperatures $\Delta T = [T_N(\infty) - T_N(d)]/T_N(\infty)$ versus the diameters, $d$, of the particles. The straight line is plotted according to the fitting result by Eq. (1).